\documentclass[11pt]{article}

\usepackage{amsmath}
\usepackage{amsfonts}
\usepackage{amssymb}
\usepackage{amscd}
\usepackage{latexsym}
\usepackage{mathrsfs}
\usepackage{amsthm}

\newtheorem{Lemma}{Lemma}[section]

\newtheorem{Theorem}[Lemma]{Theorem}
\newtheorem{Proposition}[Lemma]{Proposition}

\newtheorem{Remark}[Lemma]{Remark}

\def\og{\leavevmode\raise.3ex\hbox{$\scriptscriptstyle\langle\!\langle$~}}
\def\fg{\leavevmode\raise.3ex\hbox{~$\!\scriptscriptstyle\,\rangle\!\rangle$}}

\def\B{\mathcal{B}}
\def\BB{\mathscr B}
\def\C{\mathbb{C}}
\def\CC{\mathscr E}
\def\CCC{\mathcal C}
\def\de{\mathrm d}
\def\D{\Delta}
\def\e{\varepsilon}
\def\EE{\mathscr C}
\def\H{\mathcal{H}}
\def\JJ{\mathscr J}
\def\K{\mathcal{K}}
\def\KK{\mathscr K}
\def\O{\Omega}
\def\Pinf{\mathcal{P}_{\!\infty}}

\def\M{\mathcal M}

\def\R{\mathbb{R}}
\def\S{\mathbb{S}}
\def\P{\mathcal P}

\def\U{\mathcal{U}}
\def\T{\mathcal{T}}
\def\v{\varphi}
\def\tr{\mbox{\rm tr}}
\def\Tr{\mbox{\rm Tr}}
\def\pe{_{\scriptscriptstyle \! \bot}} 
\def\pa{_{\scriptscriptstyle \parallel}}
\def\Rep{\pi}
\def\ind{\mbox{\rm ind}}
\def\sgn{\mbox{\rm sgn}}
\def\sh{\gamma}
\begin{document}

\title{Topological boundary maps in physics: \\
General theory and applications}

\author{Johannes Kellendonk\,~and\,~Serge Richard}
  \date{\small
    \begin{quote}
      \emph{
    \begin{itemize}
    \item[]
                        Institut Camille Jordan,
                        B\^atiment Braconnier,
                        Universit\'e Claude Bernard Lyon 1, \\
                        43 avenue du 11 novembre 1918,
                        69622 Villeurbanne cedex, France
    \item[]
      \emph{E-mails\:\!:}
      kellendonk@math.univ-lyon1.fr\,~and\,~srichard@math.univ-lyon1.fr
    \end{itemize}
      }
    \end{quote}
    May 2006
  }

\maketitle

\begin{abstract}
The material presented here covers two talks given by the authors
at the conference Operator Algebras and Mathematical Physics organised in 
Bucharest in August 2005. The first one was a review given by J. Kellendonk 
on the relation between bulk and boundary topological invariants in physical 
systems. In the second talk S. Richard described an application of these ideas to 
scattering theory. It leads to a topological version of the so-called 
Levinson's theorem.
\end{abstract}
 
\section*{Introduction}
The natural language for quantum physics is 
linear operators on Hilbert spaces and underlying operator algebras. 
These algebras are fundamentally non-commutative.
Topological properties of quantum systems should hence be connected with the
topology of these algebras, which is what one calls non-commutative
topology. An important first question to be answered is therefore: what is
the correct operator algebra related to a physical system? Since we are
looking for topological effects this algebra should be a separable
$C^*$-algebra and a good starting point is to look for the  
$C^*$-version of the observables algebra. 
Once the question about the right algebra is settled we are interested
in studying its invariants asking above all: which of them have a
physical interpretation? Finally, when we have identified the
invariants, we want to derive relations between them, typically equations
between topological quantised transport coefficients or, as in
Levinson's theorem, between invariants of the bounded part and
the scattering part of the physical system.
Such relations can be obtained from topological boundary maps which do not
exist on the algebraic level.

The purpose of this paper is twofold: Explain with more details the general
theory outlined in the previous paragraph, and show its relevance in various 
applications in mathematical physics. The first section is devoted to a brief 
introduction to the natural framework of topological boundary maps and to the
description of the the general theory. The second section contains examples
of applications to solid states physics, while the third one is entirely 
dedicated to an application to potential scattering. Since crossed product
$C^*$-algebras and their twisted versions play an important r\^ole in the applications,
we have decided to incorporate an appendix on these algebras. Let us finally mention that 
Proposition \ref{itt} on the decomposition of magnetic twisted crossed product $C^*$-algebras 
as iterated twisted crossed products is of independent interest.   

\section{General theory of topological boundary maps in physics}

A $C^*$-algebra is a special kind of Banach algebra. For
our purposes, the fact that its norm satisfies the so-called $C^*$-condition
does not play an important r\^ole, but we wish it to be separable, 
{\it i.e.}~to contain a countable dense set. Infinite dimensional
von Neumann algebras are non-separable $C^*$-algebras and 
therefore not suited. 
$K$-groups are topological invariants of $C^*$-algebras: 
they are abelian groups, which are countable for separable $C^*$-algebras.
They are isomorphic for isomorphic algebras and one may think of them as
simpler objects which might tell apart $C^*$-algebras. 
A very concise formulation of our philosophy is the following: If the
$C^*$-algebra is somewhat naturally assigned to a physical system, then 
the elements of its
$K$-groups are to be understood as topological invariants of that system.

\subsection{$K$-groups and $n$-traces}

The $K_0$-group of a unital $C^*$-algebra $\EE$ is constructed from
the homotopy classes 
of projections in the set of square matrices with entries in $\EE$. 
Its addition is induced from the addition of two orthogonal projections: 
if $p$ and $q$ are orthogonal projections, {\it i.e.}~$pq=0$, then also $p+q$ is a 
projection. Thus, the sum of two homotopy classes $[p]_0+[q]_0$ is
defined as the class of the
sum of the block matrices $[p \oplus q]_0$ on the diagonal. This new class
does not depend on the choice of the representatives $p$ and $q$.
$K_0(\EE)$ is defined as the Grothendieck group of 
this set of homotopy classes of projections endowed with the mentioned addition.
In other words, the elements of the $K_0$-group are 
given by formal differences:
$[p]_0-[q]_0$ is identified with $[p']_0-[q']_0$ if there exists a projection $r$ 
such that $[p]_0+[q']_0+[r]_0 = [p']_0+[q]_0+[r]_0$. 
In the general non-unital case the
construction is a little bit more subtle. 

The $K_1$-group of a $C^*$-algebra $\EE$ is constructed from the homotopy classes 
of unitaries in the set of square matrices with entries in the unitisation of $\EE$.  
Its addition is again defined by: $[u]_1+[v]_1 = [u\oplus v]_1$ as a block matrix on the 
diagonal. The homotopy class of the added identity is the neutral element.

For our purpose, higher traces will always be constructed from ordinary
traces and derivations, which might both be unbounded. More precisely, an $n$-trace on
a $C^*$-algebra $\EE$ is determined by the data $(\T;\delta_1,\dots,\delta_n)$ where
$\T$ is a trace on $\EE$ and $\{\delta_j\}_{j=1}^n$ are 
$n$ commuting derivations on $\EE$ which leave the trace invariant:
$\T\circ\delta_j=0$. 
More important than the fact that their characters define unbounded
cyclic cocycles is for us 
that they define additive functionals on the $K$-groups by Connes'
pairing: Extending $\T$ and $\delta_j$ to matrices with
entries in $\EE$ in the canonical way one has, up to constants, 
for even $n$ a functional
$\langle{(\T;\delta_1,\dots,\delta_n)},\cdot \rangle:K_0(\EE)\to \C$ defined by
\begin{equation*}
\langle{(\T;\delta_1,\dots,\delta_n)},{[p]}\rangle =  
\sum_{\pi\in S_n}\sgn(\pi)\; \T\big(p\delta_{\pi(1)}(p)\cdots\delta_{\pi(n)}(p)\big),
\end{equation*}
and for odd $n$ a functional
$\langle{(\T;\delta_1,\dots,\delta_n)},\cdot \rangle:K_1(\EE)\to \C$ defined by
\begin{equation*}
\langle{(\T;\delta_1,\dots,\delta_n)},{[u]}\rangle = 
\sum_{\pi\in S_n}\sgn(\pi)\; 
\T\big((u^*-1)\delta_{\pi(1)}(u)\delta_{\pi(2)}(u^*)\cdots\delta_{\pi(n)}(u)\big).
\end{equation*}
Here, $S_n$ is the group of permutations of $n$ elements.

\subsection{The general theory}\label{latheogen}

Let us consider a quantum system described by a linear operator in a Hilbert space 
$\H$, and let $\EE$ be a $C^*$-subalgebra of $\B(\H)$ that is related with this system.
Here and in the sequel, $\B(\H)$ denotes the $C^*$-algebra of all bounded operators in $\H$
and $\K(\H)$ the ideal of compact operators in $\H$.
For instance, the system is described by a self-adjoint operator $H$ in $\H$
and $\EE$ contains the $C_0$-functional calculus of $H$.
Suppose now that we can identify certain elements of the $K$-groups of $\EE$
with physically meaningful quantities. For example, the spectral
projection $P_{(-\infty,c)}(H)$ of $H$ would give rise to an element of $K_0(\EE)$, 
provided $H$ is bounded from below and the value $c$ lies in a gap of the spectrum of $H$. 
Since the elements of the $K$-groups
exhibit some homotopy invariance, we expect that they
will be stable under certain perturbations of the system.
Suppose moreover that we have a higher trace such that its pairing with the
$K$-groups describes a physically significant quantity. Then this quantity is 
topologically quantised, {\it i.e.}~it takes values in a countable subgroup of the real numbers.

Now assume that we have two quantum systems, the first one related with a $C^*$-algebra
$\JJ$ and the second one with a $C^*$-algebra $\EE$. Assume moreover that these are
related via an extension, {\it i.e.}~there exists a third algebra $\CC$ such
that $\JJ$ is an ideal of $\CC$ and $\EE$ is isomorphic to the quotient
$\CC/\JJ$. Another way of saying this is that $\JJ$ and $\EE$ are the
left and right part of an exact sequence of $C^*$-algebras
\begin{equation*}
0\to \JJ\stackrel{i}{\to} \CC \stackrel{q}{\to} \EE\to 0,
\end{equation*}
$i$ being an injective morphism and $q$ a surjective morphism satisfying $\hbox{ker}\;\!
q = \hbox{im}\;\! i$.
There might not be any reasonable algebra morphism between $\JJ$
and $\EE$ but algebraic topology provides us with homomorphisms between their
$K$-groups: $\ind:K_1(\EE)\to K_0(\JJ)$ and $\exp:K_0(\EE)\to
K_1(\JJ)$, the index map and the exponential map. These maps, which
are also referred to as boundary maps, allow us to relate topological
invariants of the two systems. Furthermore, with a little luck
we also obtain dual maps on the functionals defined by
higher traces and therefore equations between numerical
topological invariants. Therefore, the procedure goes as follows:
\begin{enumerate}
\item Find a suitable $C^*$-algebra related with a given quantum system.
\item Identify $K$-elements and higher traces whose pairings 
allow for a physical interpretation.
\item Construct extensions to the $C^*$-algebras related with two different systems 
and compute the boundary maps to obtain relations between topological quantised 
quantities. 
\end{enumerate}
The reader may wonder at this point that operators don't seem to come
up at all in the picture. This is missleading. An important part of the first
and third step is actually to prove that the operators describing the
system are related in some sense to the algebra. This affiliation is often a difficult 
analytical problem! Furthermore, the
physical interpretation of pairings involves as well the operators.

\section{Applications to solid states physics}

One of the most common realisations of the ideas presented above 
is furnished by 
a quantum system described by a self-adjoint operator $H$ in a Hilbert 
space $\H$ and a norm closed subalgebra $\EE$ of $\B(\H)$ that can be
considered as the algebra of observables. In particular, $\EE$ is
expected to contain the energy observables 
$\eta(H)$, obtained from functions $\eta$ which belong to $C_0(\R)$, the algebra 
of continuous functions on $\R$ that 
vanish at infinity. This section is devoted to a presentation
of such realisations in the context of solid states physics.

Since crossed product $C^*$-algebras and their twisted versions are discussed in the
appendix, we shall not recall their definitions in this section. 

\subsection{Algebras of energy observables derived from the set of
  atomic postions} \label{sect-2.1}

An important class of $C^*$-algebras of observables can be obtained from
the geometry of the set $\P$ of equilibrium atomic positions in a solid.
$\P$ is a discrete subset of $\R^n$ which we suppose to be of
finite local complexity,
{\it i.e.}~for each $r>0$, there exists only finitely
many so-called $r$-patches $(\P-x)\cap B_r$, with $x$ varying in
$\P$. 
$B_r$ denotes the closed ball centered at $0$ and of radius $r$.

A continuous function $f: \R^n \to \C$ is called $\P$-equivariant with 
range $r$ if $B_r\cap (\P-x) = B_r\cap (\P-y)$ implies $f(x) = f(y)$,
$x,y\in \R^n$. 
The sup-norm closure of all $\P$-equivariant functions with arbitrary
range is called 
{the \it $C^*$-algebra $C_\P(\R^n)$ of $\P$-equivariant functions}. 
A typical example of a $\P$-equivariant function is a potential $V$ defined by
$V(x):=\sum_{y\in \P} v(x-y)$ where $v$ is a short range atomic potential,
{\it i.e.}~a function which decays sufficiently fast for the sum to be finite.
The unital algebra $C_\P(\R^n)$ carries the continuous $\R^n$-action $\alpha$ by translation. 
Then the algebra of the aperiodic structure described by $\P$ is
the corresponding crossed product $C^*$-algebra $\EE_\P := C_\P(\R^n)\rtimes_\alpha \R^n$.

Suppose now that the system is in an exterior constant magnetic
field whose components
we denote by $\{B_{jk}\}_{j,k=1}^n$ with $B_{jk} \in \R$. For any $x,y \in \R^n$, 
let $\omega^B(x,y):=\exp\big(-i\Gamma^B\langle 0,x,x+y\rangle\big)$, where 
$\Gamma^B\langle 0,x,x+y\rangle$ is the flux of the magnetic field though the triangle 
defined by the points $0$, $x$ and $x+y$. We refer to the appendix for the general 
construction in the case of a non-constant magnetic field with components in $C_\P(\R^n)$. 
One may thus form the magnetic twisted crossed product $C^*$-algebra 
$C_\P(\R^n)\rtimes_\alpha^B \R^n$ associated with the twisted 
actions $(\alpha, \omega^B)$ . This algebra is simply denoted by $\EE_\P^B$, and if the magnetic 
field vanishes, then this algebra corresponds to $\EE_\P$.

The important fact is the following: Let $A:=\{A_j\}_{j=1}^n$ with $A_j:\R^n \to \R$ be 
a continuous vector potential for the magnetic field. Let $H=(P-A)^2 + V$ be the Landau 
operator perturbed by a potential $V$ which is a $\P$-equivariant real function. This 
magnetic Schr\"odinger operator, which is a self-adjoint operator in $\H:=L^2(\R^n)$, is 
affiliated to $\EE_\P$ in the following sense. There exists a faithfull representation
$\Rep : \EE_\P^B\to \B(\H)$ such that for any $\eta \in C_0(\R)$, 
$\eta(H) \in \Rep(\EE_\P^B)$. Note that the representation 
is constructed with the help of the vector potential $A$. 
It can be argued that $\EE_\P^B$ is the algebra of   observables 
for the system describing the motion of an electric particle in the aperiodic solid
described by $\P$ under the influence of the constant magnetic field $B$.

Let us discuss now some examples or related constructions.

\paragraph{\em Finite systems without magnetic field.} 
Suppose that we have a finite system like an atom or a molecule and no external 
magnetic field. In that case 
$\P$ would simply be a finite set contained in some ball
$B_{r'}$, say. 
Therefore a $\P$-equivariant function with range $r$ would be constant
outside $B_{r+r'}$. Hence $C_\P(\R^n)$ is the unitisation of
$C_0(\R^n)$ and $\EE_\P = C_0(\R^n)\rtimes_\alpha \R^n + \C\rtimes\R^n$, which is
commonly called \emph{the two-body algebra}.
Let us note that the first summand is isomorphic to
$\K\big(L^2(\R^n)\big)$ and hence an ideal.
 
\paragraph{\em Crystals and quasi-crystals without magnetic field.} 
Before the discovery of quasi-crys\-tals a crystal was considered to be
a periodic 
arrangement of atoms, the set $\P$ being therefore a regular lattice. 
In that case, $C_\P(\R^n)$ is simply the algebra of continuous $\P$-periodic
functions on $\R^n$. The corresponding crossed product algebra $\EE_\P$
can be seen as the $C^*$-algebra of observables associated with this
periodic crystal. 

Idealised quasi-crystals are often described by quasi-periodic sets $\P$. 
For example such a set can be obtained from a cut \& project scheme. More
generally it has been proposed to describe aperiodic ordered systems by
repetitive Delone sets $\P$ of finite
local complexity with uniform existence of patch frequencies. 
In this case $C_\P(\R^n)$ is a lot more complicated. 
Its spectrum $\Omega_\P$ is a foliated space which
is transversally totally disconnected. Repetivity corresponds to simplicity 
of the algebra $\EE_\P$, {\it i.e.}~absense of 
non-trivial closed ideals, and uniform existence 
of patch fre\-quen\-cies to the fact that $C_\P(\R^n)$ carries a unique invariant 
normalised trace $\tau$. 
We get then a $0$-trace on $\EE_\P$ by defining
$\T(F)=\tau(F(0))$ on any continuous element $F \in L^1\big(\R^n, C_\P(\R^n)\big)$
of the crossed product. 

We add the remark that the unique invariant 
trace on $C_\P(\R^n)$ corresponds to a unique invariant ergodic
probability measure on its spectrum
$\Omega_\P$. The condition of $\P$ having uniform existence of patch
frequencies can be relaxed leading to the freedom of choice for the
trace which corresponds to a choice of ergodic probability measure on
$\Omega_\P$ and may be interpreted as a choice of physical phase.

\paragraph{\em Solids with boundary in a constant magnetic field.}
We consider a solid described by $\P$ restricted to the half-space 
$\R^{n-1}\times (-\infty,s]$, 
and therefore with a boundary at $x\pe\equiv x_n=s$. The new
variable $s$ describes the relative position between the
$\P$-equivariant potential and the boundary. We let it vary over 
$\R\cup\{+\infty\}$, $s=+\infty$ corresponding to the system without boundary.

In order to construct a suitable $C^*$-algebra for that system, we
rewrite $\EE_\P^B$ as a crossed product by $\R$,
\begin{equation}\label{gnarf}
C_\P(\R^n)\rtimes^B_\alpha\!\R^n \cong
\big(C_\P(\R^n)\rtimes^B_{\alpha_\|}\R^{n-1}\big)
\rtimes_\beta\!\R =: \EE\;.
\end{equation}
Here $C_\P(\R^n)\rtimes^B_{\alpha_\|}\R^{n-1}$ is obtained by
restricting the action to translations which are parallel to the
boundary and the twisting cocycle to 
$\omega^B\pa(x,y):=\omega^B\big((x,0),(y,0)\big)$ for $x,y\in
\R^{n-1}$. In the case of constant magnetic field
%, which is what is considered here, 
this could be understood as a choice of gauge, but a
more elegant approach which uses only gauge invariant quantities and
generalises to variable magnetic field is presented in the appendix, 
Proposition \ref{itt}. The action $\beta$ contains, of course, the
part coming from translations perpendicular to the boundary, see the appendix.
Now, for the system with boundary it is natural to consider the algebra
\begin{equation}\label{regnarf}
\CC:=\Big(C_0(\R\cup\{+\infty\})\otimes
\big(C_\P(\R^n)\rtimes^B_{\alpha_\|}\!\R^{n-1}\big)\Big)
\rtimes_{\sh\otimes\beta}\R\ ,
\end{equation}
which is the Wiener-Hopf extension of \eqref{gnarf}. 
$\gamma$ is the translation action on $C_0(\R\cup\{+\infty\})$ which
has $+\infty$ as fixed point.
The evaluation at $+\infty$ defines a surjective morphism
from $\CC$ onto $\EE$. 

The important fact, proved for $n=2$ in \cite{KS}, is the following: 
Let $b$ be the component of the magnetic field pointing in the direction 
perpendicular to the plane, and let $H_s$ be the restriction of $H=(P_1-bQ_2)^2+P_2^2 + V$ 
to the half space $\R\times (-\infty, s]$ with Dirichlet boundary conditions. 
The family $\{H_{s}\}_{s\in \R\cup\{+\infty\}}$ is affiliated to $\CC$ in the following sense: 
For any $\eta \in C_0(\R)$, there exists $F \in \CC$ such that
$\eta(H_s) = \Rep_s(F)$, where $\Rep_s:\CC\to \B\big(L^2(\R^2)\big)$
is a representation 
induced by the evaluation map at $(s,0)$, $ev_{s,0}:C_0(\R\cup\{+\infty\})\otimes C_\P(\R^2)\to \C$.
The possibility of letting $s$ tend to infinity allows to relate continuously 
$\eta(H)=\Rep_{+\infty}(F)$ with $\eta(H_s)=\Rep_s(F)$ for any $s \in \R$. Whereas the individual
representations $\Rep_s$ are not faithful, their direct sum is faithful.

From this, it can be argued that $\CC$ is the algebra of   observables for the family 
over $s$ of systems describing the motion of an electric particle, in the aperiodic 
solid described by $\P$ and under the influence of the constant magnetic field $B$, 
that is confined to the half space $\R\times (-\infty, s]$.

\paragraph{\em The edge algebra.}
Let us describe the ideal $\JJ$ which is the kernel of the surjection $\CC\to \EE$. 
It is 
\begin{equation}\label{gloups}
\JJ := \Big(C_0(\R)\otimes\big(C_\P(\R^n)\rtimes^B_{\alpha_\|}\!\R^{n-1}\big)\Big)
\rtimes_{\sh\otimes\beta}\R\ ,
\end{equation}
which is isomorphic to $\big(C_0(\R)\rtimes_\gamma\R\big)\otimes 
\big(C_\P(\R^n)\rtimes^B_{\alpha_\|}\!\R^{n-1}\big)$. 
Its elements are thus limits of elementary
tensors $F\pe\otimes F\pa$ where $F\pe$ is a compact operator, and $F\pe$
is an element of the algebra $C_\P(\R^n)\rtimes^B_{\alpha_\|}\!\R^{n-1}$. $\JJ$ is therefore 
the algebra of observables which are localised near the boundary (or edge) 
in the loose sense of being compact in the perpendicular direction. 
We call it {\it the edge algebra}.
As for $\EE_\P$ we can construct a trace on $\JJ$ starting from the 
trace $\tau$ on $C_\P(\R^n)$, namely we define $\hat\T(F\pe\otimes F\pa) :=
\Tr(F\pe) \tau\big(F\pa(0)\big)$, with $\Tr$ the standard 
trace on compact operators, on any trace class element $F\pe$ of $C_0(\R)\rtimes_\gamma \R$ 
and any continuous element $F\pa \in L^1\big(\R^{n-1}, C_\P(\R^n)\big)$
of the crossed product. 

\subsection{Examples of pairings with physical interpretation}
Let us present some systems in which the pairing of a
$K$-element with a higher trace has a physical interpretation.

The simplest construction consists in using the $K_0$-elements of an algebra of
observables $\EE_\P^B$ defined by the spectral projections and a
trace on the algebra  
to pair with them. For example, consider the $K_0$-elements defined by the 
projection $P_{(-\infty,E_F)}(H)$ of the Hamiltonian $H$ to the energies
below the Fermi energy $E_F$, provided this value lies in a gap of
spectrum of $H$.  
Pairing it with a suitable trace yields $\mbox{\rm IDS}(E_F)$, 
the integrated density of states at the Fermi level.

In the absence of magnetic field, we mention that also the pairing of the full
$K_0$-group with the $0$-trace $\T$, the so-called gap-labelling group
has been of great interest. It describes the set of possible gap
labels of a physical system.
One can construct an element in the $K_0$-group for
each $r$-patch of $\P$.
Pairing this element with the $0$-trace $\T$ yields the
frequency of the patch. Note that the notion of frequency depends on the choice
of ergodic measure on $\Omega_\P$.
As has been proved relatively recently, these frequencies generate the 
gap-labelling group \cite{viele1, viele2, viele3}. 

The most famous example is related to the topological
quantisation of the Integer Quantum Hall Effect. One typically
finds two models used to describe the quantisation. In the
bulk model, the sample is modelled by a Hamiltonian $H$ on
$L^2(\R^2)$ which is affiliated to $C_\P(\R^2)\rtimes^B_\alpha\R^2$.
Here the Hall-conductivity $\sigma_H$, the transverse component of the
conductivity tensor, is up to a universal constant the 
pairing between the $K_0$-element determined by the spectral projection
$P_{(-\infty,E_F)}(H)$, provided the Fermi energy $E_F$ lies in a gap
of its spectrum, and the $2$-trace $(\T;\delta_1,\delta_2)$.
In the representation of the algebra in $\B\big(L^2(\R^2)\big)$
discussed above, $\T$ is the trace per unit volume and $\delta_j=i[Q_j,\cdot]$, the
commutator with the $j$-component of the position operator. We refer to 
\cite{BES} for a discussion of the tight binding case where it is also
explained that the condition that $E_F$ belongs to a gap can be
relaxed to $E_F$ belongs to a mobility gap. 

\paragraph{\em Examples of pairings related to edge states.}
The edge algebra $\JJ$ is the algebra in which we expect to find the operators
describing the physics on the boundary. We first construct
an element of its $K_1$-group.

Let $\Delta$ be an interval contained in a gap of the
spectrum of the Hamiltonian $H$. Then $P_\Delta(H_s)$
will not be $0$ for $s<\infty$, but rather the projection onto the
edge states with energy in $\Delta$.
Now, consider the bounded continuous function $u: \R \to \C$ defined
for all $t \in \R $ by 
\begin{equation*}
u(t) = 1 + \chi_\Delta(t)
\Big(\exp \big(\hbox{$\frac{-2\pi i}{|\Delta|}$} (t-\inf\Delta)\big) - 1\Big)\ , 
\end{equation*}
where $\chi_\Delta$ is the characteristic function on the interval $\Delta$.
Thus, there exists an element $U -1 \in \CC$ such that 
$\Rep_s(U-1) = u(H_s) -1$.
However, since $u(H_\infty)-1 \equiv u(H)-1 = 0$, it follows that $U$ belongs
to $\JJ \subset \CC$. Therefore $U$ defines an element of $K_1(\JJ)$.

To construct odd higher traces we make use of the
trace constructed on $\JJ$ and consider the derivations
$\delta_j$ for $j\neq n$ as above and $\partial\pe:=i[P\pe,\cdot] \equiv i[P_n,\cdot]$, 
the commutator with the infinitesimal generator of translation in position space perpendicular 
to the boundary. 

The pairings of $[U]$ with the $1$-traces $(\hat\T,\delta_j)$ for $j\neq n$, and
$(\hat\T,\partial\pe)$ have physical interpretation. First of all,
the trace $\hat\T$ of an element $F\pe\otimes F\pa$ may be interpreted
as an average, namely 
the average over the position $s$ of the boundary of  
the usual trace on $L^2(\R)$ times the trace per unit 
volume on $L^2(\R^{n-1})$ of $\Rep_s(F\pe\otimes F\pa)$.
$\hbox{$\frac{1}{2\pi}$}\langle (\hat\T,\delta_j),[U]\rangle$ is the
average of
the operator $\frac{1}{|\Delta|}P_\Delta(H_s)[Q_j,H_s]P_\Delta(H_s)$.
Since this operator is $\frac{1}{|\Delta|}$ times the 
$j$-component of the current operator restricted to the edge states,
$\sigma_j:= \hbox{$\frac{1}{2\pi}$}\langle (\hat\T,\delta_j),[U]\rangle$ 
is the $j$-direction of the 
conductivity along the boundary provided the Fermi energy lies in $\Delta$. 

Similarily $\Pi:= \hbox{$\frac{1}{2\pi}$}\langle (\hat\T,\partial\pe),[U]\rangle$ 
is the average of 
$\frac{1}{|\Delta|}P_\Delta(H_s)
\frac{\partial V}{\partial x_n}P_\Delta(H_s)$.
Since the operator $P_\Delta(H_s)\frac{\partial V}{\partial x_n}P_\Delta(H_s)$ is the 
perpendicular component of the gradient force restricted to the edge
states, $\Pi$
can be understood as the gradient pressure per unit energy on the
boundary of the system again supposing that the Fermi energy lies in
$\Delta$.   

\subsection{Relating two systems}

Now we consider two quantum systems, one with an algebra of   observables
$\JJ$ the other with an algebra of   observables $\EE$, which are
related via an extension $\CC$. It turns out that in all our examples the
extensions are Wiener-Hopf extensions determined by a continuous action $\beta$ of $\R$
on some auxiliary  $C^*$-algebra $\BB$. These are abstractly defined as follows: 
Given a $C^*$-algebra $\BB$ with a continuous $\R$-action $\beta$, the Wiener-Hopf
extension of $\BB\rtimes_\beta \R$ is the crossed product $C^*$-algebra $\CC:=
\big(C_0(\R\cup\{+\infty\})\otimes \BB\big)\rtimes_{\sh\otimes\beta}\R$.
The evaluation at $+\infty$ for $C_0(\R\cup\{+\infty\})$ gives a surjective
morphism onto $\BB\rtimes_\beta\R$, which we assumed to be equal to $\EE$. 
The kernel of this morphism is supposed to be equal to $\JJ$. One good feature of the Wiener-Hopf
extension is that the $K$-groups of $\CC$ always vanish making the boundary maps $\ind$ and
$\exp$ isomorphisms. They are in fact the inverses of the Connes-Thom
isomorphism. Another advantage is that the dual maps on functionals
defined by higher traces are simple and explicit.

\paragraph{\em Solids with boundary.}
In the context of aperiodic ordered solids with boundary we have
already seen that the algebra $\CC$ is the Wiener-Hopf extension of $\EE$ with ideal $\JJ$, 
all these algebras being defined in equations \eqref{regnarf}, \eqref{gnarf} and \eqref{gloups}. 
So let us consider the boundary map on $K$-theory $\exp:K_0(\EE)\to K_1(\JJ)$.
Under the assumption that the Fermi energy $E_F$ belongs to an
interval $\Delta$ which does not overlap with the "bulk" spectrum one can show
that the image of the $K_0$-class defined by the
projection $P_{(-\infty,E_F)}(H)$ under this map is
the $K_1$-class defined by the unitary $U$ above.

To describe the dual map we consider for simplicity the case $n=2$.
In this case the map identifies the functional defined by the $2$-trace
$(\T,\delta_1,\delta_2)$ with the functional 
defined by the $1$-trace $(\hat\T,\delta_1)$. This leads to the relation
\begin{equation*}
\sigma_1 = \sigma_H 
\end{equation*}
which expresses the fact that the Hall conductivity defined as the transverse 
component of the conductivity tensor in the bulk equals the 
conductivity $\sigma_1$ of the current along the edge \cite{KRS,Johannes}. 

The dual map identifies furthermore the functional defined by the 
$0$-trace $\T$ with that defined by the $1$-trace $(\hat\T,\partial_s)$ 
where $\partial_s$ is the infinitesimal generator of translation of
the boundary. The latter functional is a linear combination of the
functionals defined by $(\hat\T,\delta_1)$ and $(\hat\T,\partial\pe)$.
As a consequence we get the relation
\begin{equation*}
\mbox{\rm IDS} = \Pi + B\sigma_1.
\end{equation*}
This relation is valid at the Fermi energy provided it belongs to a
gap of the spectrum. We note that $\mbox{\rm IDS}$ is a bulk quantity
which cannot be obtained from a measurement of the density of states
near the Fermi energy, since $\mbox{\rm IDS}(E_F)$ depends on the
density of states at all energies below $E_F$. By constrast,  $\Pi$ and
$B\sigma_1$ need only to be measured in an arbitrarily small
interval containing $E_F$. Note that they are a priori introduced as quantities 
depending on an interval containing $E_F$ but turn out to 
be largely independent of that choice.
We mention that derivating the above relation w.r.t.\ the magnetic
field strength $B$ yields Streda's formula 
$\frac{\partial}{\partial B}\mbox{\rm  IDS} = \sigma_H$ \cite{Str}.

\section{Application to scattering theory} 

Let us start by recalling very heuristically the main idea of
potential scattering.
We consider a wave packet that is prepared in the far past far enough
from a probe. 
Since we assume that the probe is of finite size, this initial wave
packet is presumably  asymptotically free.  
It is then supposed to evolve in time under the influence
of the potential describing the probe to then move far away from the probe so
that it can again be considered asymptotically free in the far future.
It is commonly expected that all the observable information of the
scattering process is 
contained in the so-called $S$-operator, an operator that relates the
initial wave  packet with the final wave packet. 
Under some weak hypotheses, this operator is unitary. 
On the other hand, the probe can possibly bind some states.
In that situation, the projection on these states is... a projection! 
Thus, we face
a situation in which there exist a unitary operator and a projection that are
related with two connected systems: 
The system of a scattering process by a probe and the
system consisting of the bound states of that probe. 
Having in mind the general theory presented in
Section \ref{latheogen}, one is naturally led to consider an algebraic 
framework that can link these two objects. 
This section is devoted to such a construction in the case
of a two-body Schr\"odinger operator. 
Other applications and extensions are in preparation \cite{KR}. 

\subsection{The framework}\label{secintro}

Let us consider the self-adjoint operators $H_0:= -\D$ and $H:= H_0 + V$
in the Hilbert space $\H := L^2(\R^n)$, where $|V(x)| \leq
c\;\!(1+|x|)^{-\beta}$  
with $\beta> 1$.
It is well known that for such short range potentials $V$, the wave operators 
\begin{equation}\label{oponde}
\O_{\pm}:=s-\lim_{t \to \pm \infty} e^{itH}\;\!e^{-itH_0}
\end{equation}
exist and have same range.
The complement of this range is spanned by the eigenvectors of $H$,
we let $P$ denote the projection on this subspace. 
The scattering operator $S$ for this system is defined by the product
$\O_+^* \O_-$, where $\O_+^*$ is the adjoint of $\O_+$. 

Levinson's theorem establishes a relation between an expression in terms of the 
unitary operator $S$ and an expression depending on the projection $P$.
There exist many presentations of this theorem, but we recall only the one 
of \cite{Martin}. 
We refer to \cite{Osborn}, \cite{Dreyfus} and \cite{Newton} for other versions 
of a similar result.

Let $\U : \H \to L^2\big(\R_+; L^2(\S^{n-1})\big)$ be the unitary
transformation that diagonalizes $H_0$, {\it i.e.}~that satisfies
$[\U H_0 f](\lambda, \omega) = \lambda [\U f](\lambda,\omega)$, with 
$f$ in the domain of $H_0$, $\lambda \in \R_+$ and $\omega \in \S^{n-1}$.
Since the operator $S$ commutes with $H_0$, there exists a family 
$\{S(\lambda)\}_{\lambda \in \R_+}$ of unitary operators in $L^2(\S^{n-1})$
satisfying  $\U S\;\! \U^* = \{S(\lambda)\}$ almost everywhere 
in $\lambda$ \cite[Chapter 5.7]{AJS}.
Under suitable hypotheses on $V$ \cite{Martin} and in the case $n=3$, 
Levinson's theorem takes the form 
\begin{equation}\label{LevMartin}
\int_0^\infty \de \lambda \;\big\{\tr[iS(\lambda)^*\hbox{$\frac{\de
    S}{\de \lambda} 
(\lambda)$}]- \hbox{$\frac{\nu}{\sqrt{\lambda}}$}\big\} = 2\pi \;\!\Tr[P],
\end{equation}
where $\tr$ is the trace on $L^2(\S^{n-1})$, 
$\Tr$ the trace on $\H$ and $\nu=(4\pi)^{-1}\int_{\R^3} \de x\;\!V(x)$.
Clearly the r.h.s.\ of this equality is invariant 
under variations of $V$ that do not change the number of bound 
states of $H$.
But it is not at all clear how this stability comes about in the 
l.h.s.

In the sequel we propose a modification of the l.h.s.\ of \eqref{LevMartin}
in order to restore the topological nature of this equality. The idea
is very natural from the point of view presented in Section \ref{latheogen}: we
rewrite the l.h.s.\ of \eqref{LevMartin} as the result of a pairing
between an element of $K_1$ and a $1$-trace.  
Beyond formula \eqref{LevMartin}, we show that the unitary $S$ is related 
to the projection $P$ at the level of $K$-theory by the index map,
{\it cf.}~Theorem \ref{thm1}. 
Let us point out that the wave operators play a key role in this work.
Sufficient conditions on $\O_-$ imply that $H$ has only a finite set of
bound states, but also give information on the behaviour of $S(\cdot)$
at the origin.

\subsection{A suitable short exact sequence and its representation}

In this section we construct a short exact sequence, {\it i.e.}~an
extension of two algebras. One algebra is associated with the
scattering system and the other with the bound state system. 
We permit ourselves to do
that twice, first in a heuristic way similar to Section~\ref{sect-2.1} 
and then again more rigorously,
shifting attention to the wave operator. This will lead us to natural
hypotheses under which we obtain a relation between 
the scattering operator $S$ and the 
projection $P$ on the bound states via a boundary map of $K$-theory.

The size of the probe being finite, it could be
described by finite set $\P$ in the spirit of Section~\ref{sect-2.1}.
For such a system we obtained the two body algebra
$\EE_\P=C_0(\R^n)\rtimes_\alpha \R^n +\C\rtimes\R^n$. 
The first summand is isomorphic to $\K\big(L^2(\R^n)\big)$, and we
expect the projection onto the bound states to be compact, supposing
that there are only finitely many. So the algebra of the bound state
system should be that ideal of $\EE_\P$. 
Although $\EE_\P$ is actually an extension of $\C\rtimes\R^n$ by the
ideal, it is not this extension which will be used.

The algebra describing the scattering part should contain all possible
$S$-operators.
Writing an $S$-operator as a unitary operator valued function 
of energy as above, it is therefore contained in
$L^{\infty}\big(\R_+; \B\big(L^2(\S^{n-1})\big)\big)$.
We now make assumptions which allow us to obtain topological information:
(1) the map $\lambda\mapsto S(\lambda)$ is continuous w.r.t.\ norm topology, (2)
$S(\lambda)-1\in \K\big(L^2(\S^{n-1})\big)$, (3) $S(0)=S(\infty) = 1$.
These will be a consequence of our hypotheses below.
They allow us to regard $S-1$ as an element of 
$C_0\big(\R_+; \K\big(L^2(\S^{n-1})\big)\big) \cong \K\big(L^2(\S^{n-1})\big)\rtimes\R$ which
we therefore consider as the algebra of the scattering system.
The action in this crossed product is trivial and the isomorphism is
given by Fourier transformation.

The extension we will use is
\begin{equation}\label{sexact}
0 \to C_0\big(\R;\KK\big)\!\rtimes_\gamma\!\R\to 
C_0\big(\R\cup\{+\infty\};\KK\big)\!\rtimes_\gamma\!\R
\stackrel{ev_{\infty}}{\to} \KK\!\!\rtimes\!\R \to 0,
\end{equation}
where $\KK$ is the algebra 
of compact operators in some Hilbert space. 
The sequence \eqref{sexact} is the Wiener-Hopf extension of the 
crossed product $\KK\!\!\rtimes\!\R$ with trivial $\R$-action on
$\KK$; $\gamma$ is the action on
$C_0\big(\R\cup\{+\infty\}\big)$  
by translation, leaving the point $\{+\infty\}$ invariant, and the
surjection $ev_{\infty}$ is induced by evaluation at $\{+\infty\}$.
Note that setting $\KK=\K\big(L^2(\S^{n-1})\big)$ we have naturally
$C_0\big(\R;\KK\big)\!\rtimes_\gamma\!\R\cong \K\big(L^2(\R^n)\big)$ and so we
expect $P$ to be in the left algebra and $S-1$ to be in the right
algebra. But instead of verifying that directly we change
perspective and concentrate on the middle algebra with the goal to
identify the wave operator as an element of it. 
To do so we represent the above
short exact sequence in the physical Hilbert space $\H$. 

Following the developments of \cite{Georgescu} we first
consider the case  $\KK=\C$ and let $A,B$ be self-adjoint
operators in $\H$ both with purely absolutely continuous spectrum
equal to $\R$ 
and commutator given formally by $[iA,B]=-1$. We can then represent 
$C_0\big(\R\cup\{+\infty\};\KK\big)\!\rtimes_\gamma\!\R$
faithfully as 
the norm closure $\CC'$ in $\B(\H)$ of the set of finite sums of the form
$\v_1(A) \;\!\eta_1(B)  + \ldots + \v_m(A) \;\!\eta_m(B)$ 
where $\v_i\in C_0\big(\R\cup\{+\infty\}\big)$ and
$\eta_i\in C_0(\R)$. We denote by $\JJ'$ the ideal obtained by
choosing functions $\v_i$ that vanish at $\{+\infty\}$.
Furthermore, we can represent $\KK\!\!\rtimes\!\R$ faithfully
in $\B(\H)$ by elements of the form $\eta(B)$ with $\eta\in C_0(\R)$. 
This algebra is denoted by $\EE'$. 

In \cite{Georgescu} position and momentum operators were chosen for
$A$ and $B$ but we take $A := -\hbox{$\frac{i}{2}$}(Q\cdot \nabla
+ \nabla \cdot Q)$ and $B:= \hbox{$\frac{1}{2}$}\ln H_0$.
We refer to \cite{Jensen} for a thorough description of $A$ 
in various representations.
Let us notice that a typical element of $\CC'$ is of the form
$\v(A)\;\!\eta(H_0)$ with $\v \in C_0\big(\R\cup\{+\infty\}\big)$
and $\eta \in C_0(\R_+)$, the algebra of continuous functions on $\R_+$ 
that vanish at the origin and at infinity.
We shall now consider $\KK=\K\big(L^2(\S^{n-1})\big)$ from the decomposition 
$\H \cong L^2\big(\R_+;L^2(\S^{n-1})\big)$ in spherical coordinates. 
Since $A$ and $H_0$ are rotation invariant the presence of a larger $\KK$ 
does not interfere with the above argument.
Thus we set $\CC:= \CC'\otimes\KK$, $\JJ:=\CC'\otimes \KK$ and
$\EE:=\EE'\otimes \KK$. These algebras are all
represented in $\B(\H)$, although $\EE$ is a quotient of $\CC$. 
The surjection $ev_{\infty}$ becomes the map
$\Pinf$, where
$\Pinf[T]:= T_\infty$, with $T_\infty$ uniquely defined by 
the conditions $\|\chi(A\geq t)\;\!(T-T_\infty)\|\to 0$ and 
$\|\chi(A\geq t)\;\!(T^*-T^*_\infty)\|\to 0$ as $t \to +\infty$, 
$\chi$ denoting the characteristic function.
We easily observe that $\Pinf[\v(A)\;\!\eta(H_0)] = \v(+\infty)\;\!\eta(H_0)$
for any $\v \in C_0\big(\R\cup\{+\infty\}\big)$ and 
$\eta \in C_0\big(\R_+; \KK\big)$, where $\v(+\infty)$ is simply the value
of the function $\v$ at the point $\{+\infty\}$. 
Let us summarise our findings: 

\begin{Lemma}\label{repdesalg}
All three algebras of \eqref{sexact} are represented faithfully in
$\H$ by $\JJ$, $\CC$ and $\EE$. In $\B(\H)$ the surjection
$ev_{\infty}$ becomes $\Pinf$.
\end{Lemma} 

Note that $\JJ$ is equal to the set 
of compact operators in $\H$.
For suitable potentials $V$, the operator $S-1$ belongs 
to $\EE$ \cite{Jensen,JK} and $P$ is a compact operator. 
The key ingredient below is the use of $\O_-$ to make the link between 
the $K_1$-class $[S]_1$ of $S$ and the $K_0$-class $[P]_0$ of $P$.

\begin{Theorem}\label{thm1}
Assume that $\O_- -1$ belongs to $\CC$. Then $S-1$ is an element of $\EE$, 
$P$ belongs to $\JJ$ and one has at the level of $K$-theory:
\begin{equation}\label{Kegalite}
\mbox{\rm ind} [S]_1 = -\; [P]_0.
\end{equation}
\end{Theorem}

\begin{proof}
Let $T\in \CC$. Then  $T_\infty=\Pinf(T) \in \EE$ satisfies
$\|\chi(A\geq t)(T-T_\infty)\|\to 0$ as $t \to +\infty$. Equivalently, 
$\|\chi(A\geq 0)\;\![U(t) T U(t)^* -T_\infty]\| \to 0$ as $t \to +\infty$, since
$T_\infty$ commutes with $U(t):=e^{\frac{i}{2}t\ln H_0}$ for all $t \in \R$.
It is then easily observed that 
$s-\lim_{t\to +\infty}U(t)\;\!T\;\!U(t)^* = T_\infty$.
Now, if $T$ is replaced by $\O_- -1$, the operator $T_\infty$ has to be equal to
$S-1$, since $s-\lim_{t \to +\infty} U(t)\;\!\O_-\;\!U(t)^*$ is equal to $S$.
Indeed, this result directly follows from the intertwining relation of $\O_-$ 
and the invariance principle \cite[Theorem 7.1.4]{ABG}.

We thus have shown that $\O_--1$  is a preimage of $S-1$ in $\CC$. 
It is well known that $\O_- \O_-^* = 1-P$ and $\O_-^* \O_- = 1$. In
particular $\O_-$ is a partial isometry so that
$\mbox{\rm ind}[S]_1 = [ \O_- \O_-^*]_0-[\O_-^* \O_-]_0=-[P]_0$, see 
{\it e.g.}~\cite[Proposition 9.2.2]{RLL}.
\end{proof}

\begin{Remark}
{\rm It seems interesting that the condition $\O_- -1 \in \CC$ implies
the finiteness of the set of eigenvalues of $H$. Another consequence of this 
hypothesis is that $S(0)=1$, a result which is also not obvious. See
\cite[Section 5]{JK} 
for a detailed analysis of the behaviour of $S(\cdot)$ near the origin.}
\end{Remark}

It is important to express the above condition on $\O_-$ in a more
traditional way, {\it i.e.}~in terms of scattering conditions.
The following lemma is based on an alternative description of the 
$C^*$-algebra $\CC$. Its easy proof can be obtained by mimicking
some developments given in Section 3.5 of \cite{Georgescu}.
We also use the convention of that reference, that is: if a symbol
like $T^{(*)}$ appears in a relation, it means that this relation has to hold
for $T$ and for its adjoint $T^*$.

\begin{Lemma}\label{appartenance}
The operator $\O_- -1$ belongs to $\CC$ if and only if 
$S(\cdot)-1$ belongs to $C_0\big(\R_+;\KK\big)$ 
and 
the following conditions are satisfied:
\begin{enumerate}
\item[{\rm (i)}] $\lim_{\e \to 0}\|\chi(H_0\leq \e) \;\!(\O_- -1)^{(*)}\|=0$, and
$\lim_{\e \to +\infty}\|\chi(H_0\geq \e) \;\!(\O_- -1)^{(*)}\|=0$,
\item[{\rm (ii)}] $\lim_{t \to -\infty}\|\chi(A \leq t) \;\!(\O_- -1)^{(*)}\|=0$, and
$\lim_{t \to +\infty}\|\chi(A \geq t) \;\!(\O_- -S)^{(*)}\|=0$.
\end{enumerate}
\end{Lemma}
Let us note that conditions {\rm (ii)} can be rewritten as 
\begin{equation*}
\lim_{t\to -\infty}\|\chi(A \leq 0)\;\!U(t)\;\!(\O_- -1)^{(*)}\;\!U(t)^*\|=0 
\end{equation*}
and 
\begin{equation*}
\lim_{t \to +\infty}\|\chi(A \geq 0)\;\!U(t)\;\!(\O_- -S)^{(*)}\;\!U(t)^*\|=0.
\end{equation*}

\subsection{The topological version of Levinson's theorem}

In the next statement, it is required that the map 
$\R_+ \ni \lambda \mapsto S(\lambda) \in \B\big(L^2(\S^{n-1})\big)$ is 
differentiable. 
We refer for example to \cite[Theorem 3.6]{Jensen} for sufficient conditions 
on $V$ for that purpose.
Trace class conditions on $S(\lambda)-1$ for all $\lambda \in \R_+$ are common requirements
\cite{Davies}. Unfortunately, similar conditions on $S'(\lambda)$ were
much less studied in the literature. However, let us already mention that these technical 
conditions are going to be weaken in \cite{KR}.

\begin{Theorem}\label{Levnous}
Let $\O_- -1$ belong to $\CC$. Assume furthermore 
that the map
$\R_+\ni \lambda \mapsto S(\lambda) \in \B\big(L^2(\S^{n-1})\big)$ is 
differentiable, and that $\lambda\mapsto \tr [S'(\lambda)]$
belongs to $L^1\big(\R_+, \de \lambda\big)$.
Then the following equality holds:
\begin{equation}\label{new}
\int_0^\infty \de \lambda \;\tr\big[i(S(\lambda)-1)^*\;\!S'(\lambda)\big] = 
2\pi \;\!\Tr[P].
\end{equation}
\end{Theorem}

\begin{proof}
The boundary maps in $K$-theory of the exact sequence \eqref{sexact}
are the inverses of the Connes-Thom isomorphism (which here
specialises to the Bott-isomorphism as the action in the quotient is
trivial) and have a dual in
cyclic cohomology \cite{Connes},
or rather on higher traces \cite{Connes2,Johannes}, which gives rise
to an equality between pairings which we
first recall: 
$\Tr$ is a $0$-trace on
the ideal $C_0\big(\R;\KK\big)\!\rtimes_\gamma\!\R\cong
\K\big(L^2(\R)\big)\otimes \K\big(L^2(\S^{n-1})\big)$ which we factor
$\Tr=\Tr'\otimes \tr$. Then $\hat \tr:  \KK\!\!\rtimes\!\R \to \C$,
$\hat \tr [a] = \tr[a(0)]$ is a trace on the crossed product and
$(a,b) \mapsto \hat \tr [a \delta(b)]$ a $1$-trace where
$[\delta(b)](t) = it b(t)$. With these ingredients 
\begin{equation}\label{pair} 
\hat \tr [i(u-1)^* \delta(u)] = -2\pi\Tr [p]
\quad\mbox{if}\quad \mbox{\rm ind}[u]_1 = [p]_0 ,
\end{equation}
provided $u$ is a representative of its $K_1$-class $[u]_1$ on which
the $1$-trace can be evaluated. This is for instance the case if
$\delta(u)$ is $\hat\tr$-traceclass.
To apply this to our situation, in which $u$ is the unitary
represented by the scattering operator and $p$ is represented by
the projection onto the bound states, we express $\delta$ and $\hat
tr$ on $\U\EE\U^*$ where $\U$ is the unitary from Section \ref{secintro}
diagonalising $H_0$. Then $\delta$ becomes $\lambda\frac{\de}{\de\lambda}$
and $\hat\tr$ becomes $\int_{\R_+} \hbox{$\frac{\de
    \lambda}{\lambda}$} \;\!\tr$. Our 
hypothesis implies the neccessary trace class property so that
the l.h.s.~of (\ref{pair}) corresponds to 
$ \int_0^\infty \de \lambda \;\!\tr\big[i(S(\lambda)-1)^*
S'(\lambda)\big]$ and the r.h.s.~to $2\pi\;\!\Tr[P]$.
\end{proof}

\begin{Remark}
{\rm Expressions very similar to
\eqref{new} already appeared in \cite{Osborn} and \cite{Dreyfus}.
However, it seems that they did not attract the attention of the
respective authors and that a formulation closer to
\eqref{LevMartin} was preferred.  One reason is that the operator
$\{S(\lambda)^*S'(\lambda)\}_{\lambda \in \R_+}$ has a physical
meaning: it represents the {\em time delay} of the system under
consideration.  We refer to \cite{AC} for more explanations and
results on this operator.}
\end{Remark}

\begin{Remark}
{\rm At present our approach does not allow to say anything about a
{\em half-bound state} but this will be remedied in \cite{KR}. 
We refer to \cite{JK}, \cite{Newton} or
\cite{Newton2}   
for explanations on that concept and to \cite{Newton} or \cite{Newton2} for
corrections of Levinson's theorem in the presence of such a 
{\em 0-energy resonance}.}
\end{Remark}

\subsection{Further prospects}

We outline several improvements or extensions that ought to be carried
out or seem natural in view of this note. We hope to express some of these
in \cite{KR}.
\begin{enumerate}
\item 
Our main hypothesis of Theorem~\ref{Levnous}, that $\O_--1$
belongs to the $C^*$-algebra $\CC$, is crucial and we have provided
estimates in Lemma \ref{appartenance} which would guarantee it.
Such estimates are rather difficult to obtain and we were
not able to locate similar conditions in the literature. They clearly
need to be addressed.
\item Similar results should hold for a more general
operator $H_0$ with absolutely continuous spectrum.
In that case, the role of $A$ would be played by an operator
conjugate to $H_0$. We refer to \cite[Proposition 7.2.14]{ABG} for the 
construction of such an operator in a general framework.
\item
More general short range potentials or trace class perturbations can also 
be treated in a very similar way. By our initial hypothesis on $V$ we have
purposely eliminated positive eigenvalues of $H$, but it would be interesting
to have a better understanding of their role with respect to 
Theorems \ref{thm1} and \ref{Levnous}.
\item 
In principle, Theorem~\ref{thm1} is stronger than Levinson's
theorem and one could therefore expect new topological
relations from pairings with other cyclic cocycles. In the present
setting these do not yet show up as the ranks of the
$K$-groups are too small. But in more complicated scattering
processes this could well be the case.    
\item 
In the literature one finds also the
so-called {\it higher-order Levinson's Theorems} \cite{Bolle}. 
In the case $n=3$ and under suitable hypotheses they take the form
\cite[equation 3.28]{Bolle} 
\begin{equation*}
\int_0^\infty \de \lambda \;\!\lambda^{\!\hbox{\tiny \it N}}
\big\{\tr \big[iS(\lambda)^*S'(\lambda)\big] - C_{\!\hbox{\tiny \it
    N}}(\lambda)\big\} = 2\pi \sum_j e_j^{\hbox{\tiny \it N}}, 
\end{equation*}
where $N$ is any natural number, $C_{\!\hbox{\tiny \it N}}$ are correction terms, 
and $\{e_j\}$ is the set of eigenvalues of $H$ with multiplicities counted. 
The correction terms can be explicitly computed in terms of 
$H_0$ and $V$ \cite{Bolle} and we expect that they can be absorbed in
a similar manner into the $S$-operator as above. 
\end{enumerate}

\section{Appendix on twisted crossed products}

In this section, we start by recalling the definition of a twisted crossed product 
$C^*$-algebra borrowed from \cite[Section 2]{MPR}. We refer to the references quoted 
in that paper for a more general definition. Then, we consider the particular
situation of the $2$-cocycle defined by a magnetic field. Finally, a decomposition of 
the magnetic twisted crossed product as an iterated twisted crossed product is proved.   

Let $X$ be an abelian, second countable, locally compact group, and let $\CCC$ 
be an abelian $C^*$-algebra, with its norm denoted by $\|f\|$ 
and its involution by $f^*$, for any $f \in \CCC$. 
Assume that there exists a group morphism $\alpha: X \to \mathrm{Aut}(\CCC)$
from $X$ to the group of automorphisms of $\CCC$ such that the map
$X\ni x \mapsto \alpha_x[f] \in \CCC$ is continuous for all $f \in \CCC$.
Assume also that there exists a strictly continuous normalized $2$-cocycle
$\omega$ on $X$ with values in the unitary group of the multiplier algebra $\M(\CCC)$ 
of $\CCC$. We refer to \cite{WO} for the definition of the multiplier algebra, but recall
that if $\CCC \subset\B(\H)$, for some Hilbert space $\H$, then $\M(\CCC)=\{a \in \B(\H)\, |\,
af, fa \in \CCC, \forall f \in \CCC \}$. Since $\CCC$ is abelian, so is $\M(\CCC)$. 
In other words, $\omega: X \times X \to \M(\CCC)$
satisfies the following conditions: For any $x,y,z \in X$ and $f \in \CCC$, 
(1)~$\omega(x,y)^*\omega(x,y)=1$, 
(2)~the map $X\times X \ni (x,y) \mapsto \omega(x,y)f \in \CCC$ is continuous, 
(3)~$\omega(x,0) = \omega(0,x) = 1$, and 
(4)~the 2-cocycle relation
\begin{equation}\label{2cc}
\omega(x,y)\;\!\omega(x+y,z)=\alpha_x[\omega(y,z)]\;\!\omega(x,y+z)\;\! \ .
\end{equation} 
We have used in this relation that any automorphism of $\CCC$ extends uniquely to 
an automorphism of $\M(\CCC)$. 
The quadruple $(\CCC, \alpha, \omega, X)$ is usually called {\it an abelian twisted
$C^*$-dynamical system}.
Finally, we shall also assume the additional condition $\omega(x,-x)=1$, which holds
in all the applications we have in mind.

Now, let $\kappa \in [0,1]$; this additional parameter is convenient in order to relate 
our expressions with earlier results found in the literature. The special cases
$\kappa = 0$ and $\kappa=1$ are related with the right and the left quantisation respectively.
Most of the time only the case $\kappa=0$ is presented, but $\kappa=1/2$ is prefered in
quantum mechanics because of some additional symmetry properties. However, let us already
mention that the following structures are isomorphic for different $\kappa$.

We consider the set $L^1(X;\CCC)$ endowed with the norm $\|F\|_1:=
\int_X \|F(x)\|\de x$ for any $F \in L^1(X;\CCC)$, the multiplication
\begin{equation*}
(F\diamond G)(x):=\int_X\alpha_{\kappa(y-x)}[F(y)]\;\!\alpha_{(1-\kappa)y}
[G(x-y)]\;\!\alpha_{-\kappa x}[\omega(y,x-y)] \;\!\de y \ ,
\end{equation*}
and the involution $F^\diamond(x):=\alpha_{(1-2\kappa)x}[F(-x)^*]$.
Then the envelopping $C^*$-algebra of $L^1(X;\CCC)$ endowed with these operations
is called {\it the twisted crossed product of $\CCC$ by $X$ associated with the twisted 
actions $(\alpha, \omega)$}. We shall denote it by $\CCC \rtimes_\alpha^\omega X$, or
simply $\CCC\rtimes_\alpha X$ if $\omega \equiv 1$.

Let us now assume that $\CCC$ is a $C^*$-algebra of bounded and uniformly continuous
functions on $\R^n$, stable under translations. We also fix $X:= \R^n$ and the action $\alpha$
of $\R^n$ on $\CCC$ is simply given by translations.
Moreover, suppose that a continuous magnetic field on $\R^n$ is 
also present. Its components are denoted by $\{B_{jk}\}_{j,k=1}^n$, and for any 
$q, x, y \in \R^n$ we define 
\begin{equation*}
\omega^B(q;x,y):=\exp\big(-i\Gamma^B\langle q,q+x,q+x+y\rangle\big) \ ,
\end{equation*}
where $\Gamma^B\langle q,q+q,q+x+y\rangle$ is the flux of the magnetic field through 
the triangle defined by the points $q$, $q+x$ and $q+x+y$. 
If all $B_{jk}$ belong to $\CCC$, then the map 
$\omega^B:\R^n \times \R^n \ni (x,y)\mapsto \omega^B(\cdot; x,y) \in \M(\CCC)$ satisfies
all the above conditions imposed on $\omega$.
Furthermore, the additional property $\omega^B(x,-x)=1$ is always fulfilled.
One may thus form the magnetic twisted crossed product $C^*$-algebra associated with the 
magnetic twisted actions $(\alpha, \omega^B)$ and denote it simply by 
$\CCC\rtimes_\alpha^B \R^n$.

We now show how the magnetic twisted crossed product 
$\CCC\rtimes^B_\alpha \R^n$ 
can be decomposed as an iterated twisted crossed product.
The strategy is inspired from \cite[Theorem 4.1]{PR1} which 
deals with more general algebras $\CCC$, 2-cocycles $\omega$ and
groups $X$, but only in the special case $\kappa = 0$.
Let us first observe that if $(\CCC,\alpha,\omega,\R^n)$ is an
abelian twisted $C^*$-dynamical system, then 
$(\CCC,\alpha\pa,\omega\pa, \R^{n-1})$, with $\omega\pa(x,y):=
\omega\big((x,0),(y,0)\big)$ for all $x,y \in \R^{n-1}$ and
$\alpha\pa$ the restriction of the action to $\R^{n-1}$, 
is also an abelian twisted $C^*$-dynamical system.
For simplicity, we shall keep writing $\alpha$ and $\omega$ for
$\alpha\pa$ and $\omega\pa$,
and $\diamond$, ${}^\diamond$ for the multiplication and the 
involution in $L^1(\R^{n-1};\CCC)$. Furthermore, we omit the superscript
$B$ in $\omega^B$ in the following statement and in its proof.

\begin{Proposition}\label{itt}
For any magnetic abelian twisted $C^*$-dynamical system $(\CCC,\alpha,\omega,\R^n)$ 
and any $\kappa \in [0,1]$, there exits a 
continuous group morphism  $\beta$ from $\R$ to the group of 
automorphisms of $\CCC\rtimes^\omega_\alpha\!\R^{n-1}$ 
such that 
\begin{equation}\label{detwiste}
\CCC\rtimes^\omega_\alpha\!\R^n \cong
\big(\CCC\rtimes^\omega_\alpha\!\R^{n-1}\big)
\rtimes_\beta\!\R\;.
\end{equation}
\end{Proposition}

\begin{proof}
In this proof, we consider the elements $x\pa,y\pa$ of $\R^{n-1}$ and 
the elements $x\pe,y\pe$ of $\R$. For  $F \in L^1(\R^{n-1};\CCC)$, 
let us set
\begin{equation*}
\beta_{x\pe}[F](x\pa):=
\alpha_{-\kappa (x\pa,0\pe)}[\Box(x\pe,x\pa)]\; \alpha_{(0\pa,x\pe)}[F(x\pa)]
\end{equation*}
with $\Box(x\pe,x\pa):=\omega\big((0\pa,x\pe),(x\pa,0\pe)\big)\;\!
\omega\big((x\pa,0\pe),(0\pa,x\pe)\big)^*$. 
We may observe that for each $q \in \R^n$ the expression
$\Box(q;x\pe,x\pa)$ is equal to $\exp\!\big\{-i\Gamma^B_{\!\!\Box} \big(q;(0\pa,x\pe),
(x\pa,0\pe)\big)\big\}$, 
where the exponent is the flux  of the magnetic field trough the square defined by the points 
$q$, $q+(0\pa,x\pe)$, $q+(x\pa,x\pe)$ and $q+(x\pa,0\pe)$.

(i) Let us first prove that $\beta$ defines a continuous group morphism from $\R$ to 
$\mathrm{Aut}(\CCC\rtimes^\omega_\alpha\!\R^{n-1})$.
It is obvious that $\beta_{x\pe}[F]$ belongs to $L^1(\R^{n-1};\CCC)$.
By taking into account the relation \eqref{2cc} and the 
special property $\omega(x,tx)=1 \ \forall  x\in \R^n$ and $t\in \R$ 
of magnetic 2-cocycles, one also easily obtains that 
$\beta_{x\pe}\big[\beta_{y\pe}[F]\big] = \beta_{x\pe + y\pe}[F]$. 
Furthermore, for $G \in L^1(\R^{n-1};\CCC)$ one has
$\beta_{x\pe}[F\diamond G] = 
\beta_{x\pe}[F]\diamond \beta_{x\pe}[G]$ if for all $y\pa \in \R^{n-1}$ 
the following equality holds:
\begin{equation*}
\Box(x\pe,x\pa) = \Box(x\pe,y\pa)\;\alpha_{(y\pa,0\pe)}[\Box(x\pe,x\pa-y\pa)]
\;\omega(y\pa,x\pa-y\pa)\;\alpha_{(0\pa,x\pe)}
[\omega(y\pa,x\pa-y\pa)]^*. 
\end{equation*}
But again, this can be verified with the help of relation \eqref{2cc}.
The same relation also leads to the equality $\beta_{x\pe}[F^\diamond] =
(\beta_{x\pe}[F])^\diamond$. 
Finally, the continuity of the map $\R \ni x\pe \mapsto \beta_{x\pe}[F] \in L^1(\R^{n-1};\CCC)$
can be proved by taking into account the strict continuity of $\omega$ and the continuity 
of the map $\alpha: \R^n \to \mathrm{Aut}(\CCC)$. 
By a density argument, one completes the proof of the assertion. 

(ii) Let us now define the bijective map:
$C_c(\R^n;\CCC) \ni F \mapsto F' \in C_c\big(\R;C_c(\R^{n-1};\CCC)\big)$ 
given by 
\begin{equation*}
F'(x\pa;x\pe):=\alpha_{-\kappa x}\big[\Box(\kappa x\pe,x\pa)^*\;\!
\omega\big((x\pa,0\pe),(0\pa,x\pe)\big)^*\big]\;\!F(x\pa,x\pe).
\end{equation*}
The multiplication in $C_c\big(\R;C_c(\R^{n-1};\CCC)\big)$ is defined by
\begin{equation*}
(F' \diamond_\beta G')(\cdot;x\pe):=
\int_\R \de y\pe \ \beta_{\kappa(y\pe-x\pe)}[F'(\cdot; y\pe)]\;\!\diamond\;\!
\beta_{(1-\kappa)y\pe}[G'(\cdot; x\pe-y\pe)]
\end{equation*}
and the involution is given by $(F')^{\diamond_\beta}(\cdot;x\pe)=
\beta_{(1-2\kappa)x\pe}\big[\big(F'(\cdot;-x\pe)\big)^{\diamond}\big]$, 
{\it i.e.}~:
\begin{equation*}
(F')^{\diamond_\beta}(x\pa;x\pe)=
\alpha_{-\kappa (x\pa,0\pe)}\big[\Box\big((1-2\kappa)x\pe,x\pa\big)\big]\ 
\alpha_{(1-2\kappa)x}[F'(-x\pa;-x\pe)^*].
\end{equation*}
The final step consists in verifying that $F' \diamond_\beta G'$
is equal to $(F \diamond G)'$, and that 
$[(F')^{\diamond_\beta}]$ is equal to $(F^{\diamond})'$.
These equalities can be checked without difficulty by taking into account 
the relation \eqref{2cc} and the already mentioned property  
of magnetic 2-cocycles. 
A density argument completes the proof.
\end{proof}

\section*{Acknowledgements}
Serge Richard thanks the Swiss National Science Foundation and the  european network: 
Quantum Spaces - Noncommutative Geometry for financial support.

\end{document}